# Wave Manipulations by Coherent Perfect Channeling

Xiaonan Zhang, Chong Meng, and Z. Yang*

Department of Physics, the Hong Kong University of Science and Technology

Clearwater Bay, Kowloon

Hong Kong, China

**We report experimental and theoretical investigations of coherent perfect channeling (CPC), a process that two incoming coherent waves in waveguides are completely channeled into one or two other waveguides with little energy dissipation via strong coherent interaction between the two waves mediated by a deep subwavelength dimension scatterer at the common junction of the waveguides. Two such scatterers for acoustic waves are discovered, one confirmed by experiments and the other predicted by theory, and their scattering matrices are formulated. Scatterers with other CPC scattering matrices are explored, and preliminary investigations of their properties are conducted. The scattering matrix formulism makes it possible to extend the applicable domain of CPC to other scalar waves, such as electromagnetic waves and quantum wavefunctions.**

Strong interaction between two coherent waves can take place when mediated by a suitable scatterer in deep subwavelength dimension, and lead to extraordinary effects. In coherent perfect absorption (CPA) of electromagnetic waves, which was theoretically predicted as the reverse process of lasing [1], the scatterer is an absorber with suitable transmission, reflection, and absorption coefficients [1]. It was confirmed experimentally first in a silicon cavity [2], and then in dielectric metasurfaces [3]. The process in a strongly interacting system was also observed [4]. In coherent perfect rotation [5], which refers to reversible processes transferring any fixed input polarization state of coherent counter propagating light waves completely into its orthogonal polarization, the mediator is a dielectric Faraday rotator. Bending of light beams to the wrong side of the normal direction of a corrugated free-standing metal film under bidirectional coherent illumination was also predicted [6]. In CPA of acoustics waves in waveguides [7], the scatterers were made of decorated membrane resonators (DMR's) [8] and hybrid membrane resonators (HMR's) [9, 10]. The highest ratio of incoming wave intensity over the outgoing one obtained at optimum acoustic CPA conditions was 975 times, and the phase sensitivity of the process was demonstrated. In many applications, such as all-wave interference logic gates and integrated optical circuits [11 – 15], waveguide modes rather than open space ones are preferred. Plasmon-based interferometric logic operations in silver nanowire networks [11] and nanoscale plasmonic slot were demonstrated [12]. All-optical AND, XOR, and NOT logic gates based on Y-branch were also investigated in photonic crystal waveguide [13]. Likewise, acoustic logic gates and Boolean operations based on self-collimating acoustic beams [14] and density-near-zero metamaterials [15] were investigated. However, the logic operations are not 'perfect' in that there are always residue reflection in the input channels and spurious scattering waves at the gate junction.

We report first the experimental demonstration of coherent perfect channeling (CPC) of acoustic waves. In the three-port T-configuration waveguide system, in seemingly the same conditions for CPA for the scatterer at the junction where two counter propagating waves with the same phase and amplitude are incident from the two horizontal main waveguides [7], all the incident waves are actually turned, or channeled, into the vertical side branch waveguide with only 3.2 % of the wave energy truly dissipated in the process. In the reverse process, which is ensured by time reversal symmetry and confirmed by simulations, an input wave in the side branch is totally split into two coherent outputs in the two main waveguides with perfectly matched phase and amplitude. Theoretical studies also predict similar phenomenon in four-port X-configuration waveguide system, in which counter propagating waves in the horizontal waveguides are perfectly channeled into the two vertical ones, and vise versa. Scattering matrices for the three-port and four-port configurations are then formulated, which preserve time reversal symmetry and total flux. This implies that the findings in acoustic waves could be generalized to other scalar waves, such as electromagnetic waves, quantum wavefunctions, and spin waves [16], where time reversal symmetry and linear superposition principle are upheld. The potential of the CPC process in a wide range of applications, including all-wave logic gates, interferometry, coherent source arrays, and other coherent perfect manipulation of waves, are discussed.

The experimental setup for CPC is schematically shown in Fig. 1. The horizontal left and right main waveguides (port-1 and -2) are the same as the one used in earlier CPA experiments [7]. A third waveguide (side branch) with twice the cross section area as the horizontal ones is connected at the T-junction to the main waveguides. A DMR with a rubber membrane 61 mm in diameter and decorated by a platelet 5.5 mm in diameter and 30 mg in mass is mounted at the entrance of the side branch. The far end of the side branch is terminated by an anechoic sponge wedge. Viewed in the main waveguides, the side branch with the DMR serves as a monopole resonator similar to an HMR mounted on the sidewall of the main waveguide [7].

Numerical simulations using the COMSOL MultiPhysics software package were carried out to verify the underline mechanism for CPC in various configurations. Actual device structures parameters were used in the simulations when they were applicable. The mass density, Poisson's ratio, Young's modulus, and the pre-stress of the membrane were 940 $kg/m^3$, 0.49, $2 \times 10^5$, and $0.5 MPa$, respectively. The dissipation is introduced in the form of an imaginary part in the tension that is about 1 % of the real part, similar to previous works [8 – 10].

Similar to the procedure to achieve CPA, the transmission $t$ and the reflection $r$ of the junction under one side incidence (say from port-1) in the main waveguide were measured first. Figure 2(a) shows the experimental transmission (red circles) and reflection (green circles) spectra together with the simulation results (appropriately colored solid curves). At 248.3 Hz the transmission is 0.496 and the reflection is 0.498. Both are very close to the critical value of 0.5 for CPA. The phase of the reflection is 176.2° while that of the transmission is −1.3°, as shown in Fig. 2(b) by the corresponding red and green circles together with the simulation results in solid curves of appropriate colors. The seemingly lost wave energy in the main waveguides, however, was actually channeled to the side branch, as the outgoing wave amplitude in the side branch (the turning coefficient $\tau$) reached 0.49 in the meantime, as shown in Fig. 2(a) as the blue circles. Total wave energy was conserved because the side branch cross section area is twice of that of the main waveguides, i. e., $r^2+t^2+2\tau^2$ is very close to 1. In the theoretical simulations, the turning coefficient nearly coincides with the reflection because both waves are emitted by the DMR, so it is not shown in Fig. 2(a) and 2(b). The transmission and reflection seen in the main waveguides are therefore identical to those for a monopole CPA [7].

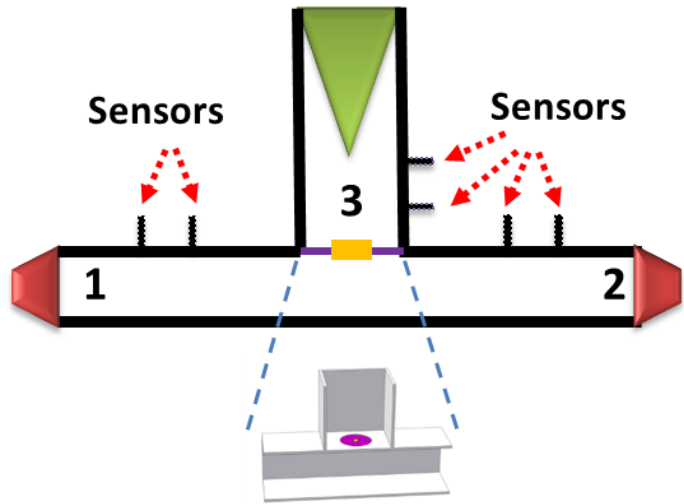

Figure 1. The schematics of the experimental setup.

Figure 2 (c) shows the apparent CPA seen in the main waveguides. At 248.2 Hz, the experimental outgoing wave amplitudes (purple circles for the left outgoing and green circles for the right outgoing waves) drop to 1.7 % and 2.3 %, respectively, of the incoming ones emitted by the left and the right speakers. The theoretical amplitudes for the left outgoing and the right outgoing waves are identical, so only one of them is shown as the solid green curve in the figure. The amplitudes of the incoming waves (not shown in the figure) remain nearly 1 in the entire frequency range of interest. The intensity ratio of the incoming waves over the outgoing ones is over 2400 times, which is significantly higher than the best CPA scatterer reported earlier [7].

What is truly different from real CPA is that the outgoing wave amplitude in the side branch reaches maximum of 0.99 at the apparent CPA frequency of 248.2 Hz. Nearly all the incoming wave energy in the main waveguides is therefore channeled into the outgoing wave in the side branch. This is further supported by the measured apparent main waveguide loss coefficient (red circles) and the turning coefficient into the side branch (green circles) shown in Fig. 2(d). The two coefficients are nearly the same, showing that the loss in the main waveguide has been mostly channeled into the outgoing wave in the side branch. At the optimum frequency, nearly 97 % of the energy is channeled into the side branch, while 3.2 % of the energy is dissipated by the DMR. The process therefore deserves the name of Coherent Perfect Channeling.

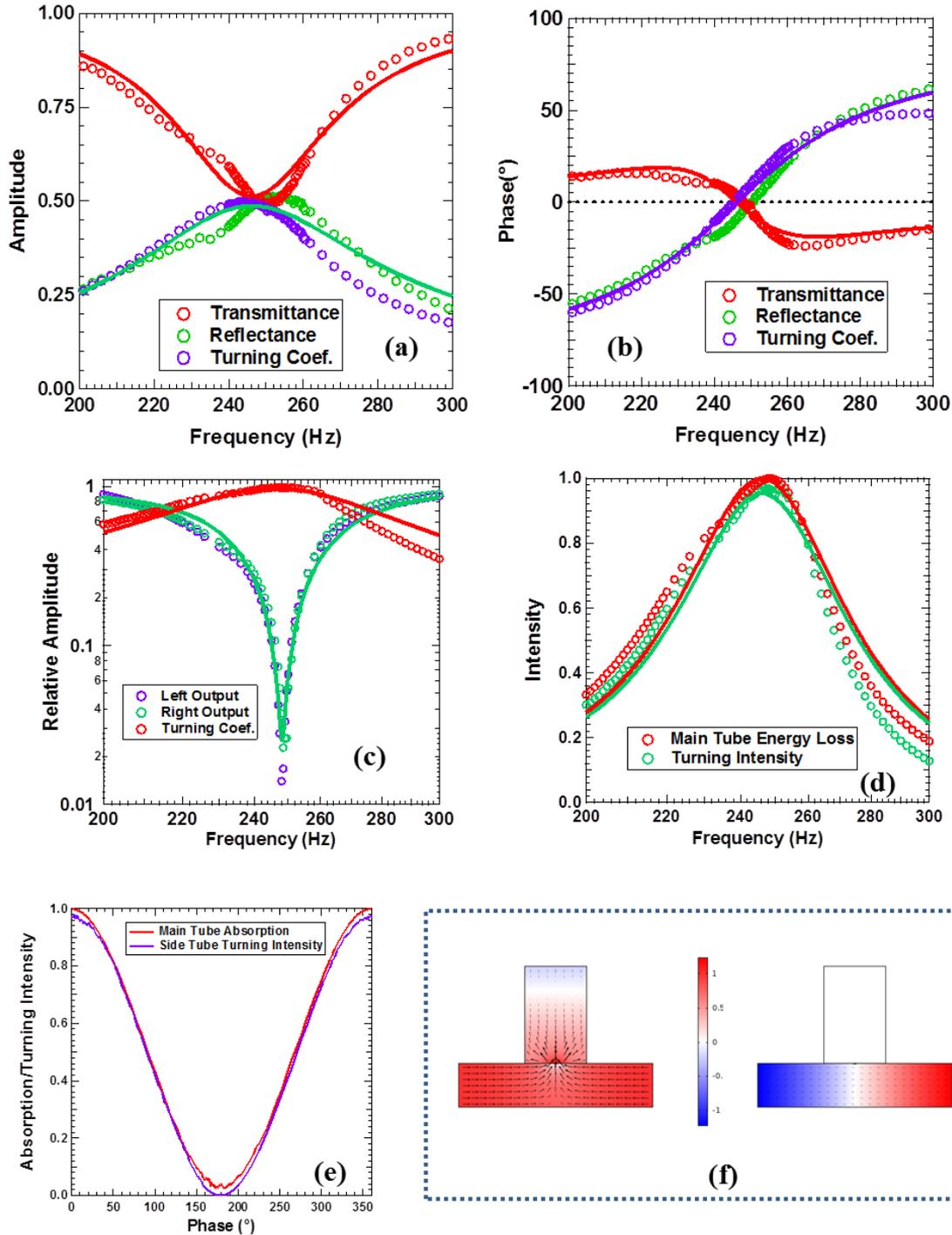

Figure 2. In (a) through (d), the circles are experimental data and the solid curves are theoretical simulation results. (a) The one-side incidence transmission (red), reflection (green) coefficients in the main waveguides and the turning coefficient (blue) from the main waveguides into the side branch. (b) The phase spectra of the transmission (red), reflection (green), and the turning coefficient (blue). For clarity, the reflection phase has been shifted downwards by 180°. (c) The CPC spectra consisting of the amplitudes of the outgoing waves (green and purple) in the main waveguides and the turning coefficient into the side branch (red). (d) The apparent CPA intensity (red) and the turning intensity (green) spectra. (e) The experimental apparent CPA intensity and the turning intensity as a function of the phase difference between the two incoming waves in the main waveguide. (f) The pressure and air velocity fields at zero phase difference (left) and 180° out of phase (right) between the two incoming waves.

Another critical feature in the two-wave coherent interplay is the dependence of the outcome, being absorption, rotation, or in this case channeling, on the relative phase $\phi$ of the incoming waves in the main waveguides. Indeed, as shown in Fig. 2(f), both the experimental main waveguide absorption and the turning coefficient exhibit $\cos^2(\phi/2)$ dependence, which is expected for monopolar CPA [7]. The minimum turning coefficient is $3.7\times10^{-5}$ reached at $\phi = 179.9°$. The contrast ratio of the turning coefficient at maximum of 0.97 over the minimum is therefore $2.6 \times 10^4$ times, or 44 dB. If used as a phase sensitive detector at this optimum phase difference, the intensity will change by nearly 7 times per degree, 26 times more sensitive than the best one reported earlier [7].

As shown in Figs. 2 (a) through (d), the theoretical predictions agree well with the general features of the experimental results. We therefore conducted theoretical studies to explore further the properties of CPC. Shown in Fig. 2(f) are the pressure and air velocity fields obtained from the simulations at optimum CPC in-phase condition (left) and out-of-phase condition (right). It is seen that at CPC the incoming waves in the main waveguides are channeled completely into the side branch. However, unlike the real CPA case for the side mounted HMR [7] in which the membrane vibration is enhanced by nearly 70 times for wave energy dissipation, here the air velocity is enhanced by about 14 times, resulting in the real dissipation of only 3.2 %. At CPC the platelet is vibration in unison with the membrane, similar to that of the first eigenmode of the DMR at 259.8 Hz. At 180° off phase, the waves mostly stay in the main waveguide with little channeling into the side branch, and the membrane remains almost motionless.

The near-zero dissipation implies that CPC obeys time reversal symmetry, which dictates that the reverse of CPC can also take place. Indeed, the simulations show that at the CPC frequency, the incoming wave in the side branch will almost totally split into two identical outgoing ones in the main waveguides, with little reflection at the junction. The frequency dependence of the reflection and the turning coefficient in the reverse CPC resemble closely the spectra in Fig. 2(c), with the reflection following the green curve (forward CPC main waveguide outgoing waves), and the turning coefficient following the red curve (forward CPC turning coefficient).

In order to conceptualize the CPC process, we analyze the process in terms of the scattering matrix of a scatterer at the junction. Consider an *N*-port waveguide system with cross section areas $S_1, S_1,..., S_N$, intersecting at a common junction where a scatterer is located. The incoming scalar waves and the outgoing ones are related by the $N \times N$ scattering matrix of the scatterer at the junction given by $(O_1 \ O_2 \ ... \ O_N)^T = \hat{M}_N (I_1 \ I_2 \ ... \ I_N)^T$. If the net wave flux is conserved in the process, then

$$\sum_{i=1}^{N} S_i I_i^2 = \sum_{i=1}^{N} S_i O_i^2 \qquad (1).$$

The process must obey time reversal symmetry as dictated by the wave equation, if net flux is conserved. In the cases where all the waveguides contain the same medium, we then must also have $(I_1 \ I_2 \ ... \ I_N)^T = \hat{M}_N (O_1 \ O_2 \ ... \ O_N)^T$. This leads to $\hat{M}_N \cdot \hat{M}_N = \hat{I}_0$ the unity matrix. The scattering matrix may not be symmetric because of the cross section area difference between different waveguides. The two-port monopole CPA process is irreversible, as the matrix $\hat{M}_2 = \frac{1}{2}\begin{pmatrix} 1 & -1 \\ -1 & 1 \end{pmatrix}$ does not satisfy the $\hat{M}_2 \cdot \hat{M}_2 = \hat{I}_0$ condition. It is expected because half of the wave energy is dissipated in the process, and time reversal symmetry is broken.

The scattering matrix for the T-configuration CPC presented above can be expressed as

$\hat{M}_3 = \begin{pmatrix} r & t & \tau' \\ t & r & \tau' \\ \tau & \tau & r' \end{pmatrix} = \frac{1}{2}\begin{pmatrix} -1 & 1 & 2 \\ 1 & -1 & 2 \\ 1 & 1 & 0 \end{pmatrix}$. One can verify by simple algebra that $\hat{M}_3 \cdot \hat{M}_3 = \begin{pmatrix} 1 & 0 & 0 \\ 0 & 1 & 0 \\ 0 & 0 & 1 \end{pmatrix}$, and for any incoming waves $(a \ b \ c)^T$, the flux of both the outgoing waves $P_O$ and incoming waves $P_I$ are equal to $a^2 + b^2 + 2c^2$. It is essential that the third channel must be twice in area of the other two, as will be shown below. The matrix implies the following. In the main waveguide connecting port-1 and port-2, the reflection to an incoming wave from port-1 or port-2 at the junction is –0.5, the transmission is 0.5, and the turning coefficient into the side branch to port-3 is also 0.5. When two identical waves are emitted from port-1 and port-2 simultaneously, that is $\hat{I} = (1 \ 1 \ 0)^T$, the outgoing waves are $\hat{O} = (0 \ 0 \ 1)^T$, which is the CPC process. For an incoming wave in the side branch from port-3, $\hat{I} = (0 \ 0 \ 1)^T$, the reflection at the junction is 0, and the turning coefficient is 1, so $\hat{O} = (1 \ 1 \ 0)^T$. This is the reverse CPC process. All these characteristics of the scattering matrix agree well with the experimental results and theoretical simulations shown in Fig. 2.

If waveguides with equal cross section area are preferred, one could use four-port systems in an X-configuration as shown in Fig. 3(a). A type of scatterers have been discovered which have the following scattering matrix

$$\hat{M}_4 = \begin{pmatrix} r & t & \tau & \tau \\ t & r & \tau & \tau \\ \tau & \tau & r' & t' \\ \tau & \tau & t' & r' \end{pmatrix} = \frac{1}{2}\begin{pmatrix} -1 & 1 & 1 & 1 \\ 1 & -1 & 1 & 1 \\ 1 & 1 & -1 & 1 \\ 1 & 1 & 1 & -1 \end{pmatrix} \quad (2)$$

which can be verified by simple algebra that it preserves total wave flux and time reversal symmetry. Such scatterers could be realized, as predicted by simulations, by mounting a DMR at the junction entrance of each secondary waveguide-3 and -4, as shown in Fig. 3(a). In the present case, the two DMR's are the same as the one for the T-configuration CPC shown in Fig. 2. At 250.8 Hz, which is slightly different from the T-configuration CPC frequency because of the change in geometry, the reflection for the incoming waves from any port is –0.5, and the transmission and the turning coefficients into the three other waveguides are 0.5. Therefore, viewed either through the horizontal waveguides (waveguide-1 and -2) or the vertical waveguides (waveguide-3 and -4), the junction looks exactly like a CPA monopole [7]. The difference is that the lost wave energy in the horizontal waveguides is channeled equally into the perpendicular waveguides, instead of being dissipated. The same is true for the perpendicular waveguides. If two identical incoming waves are launched in waveguide-1 and -2, respectively, they will be channeled completely and equally into waveguide-3 and -4, as shown in Fig. 3(a). Likewise, due to time reversal symmetry, identical incoming waves in waveguide-3 and -4 will be channeled completely and equally into waveguide-1 and -2. In fact, as dictated by the matrix in Eq. (2) and verified by simulations, identical incoming waves from any pairs of the four ports will be channeled completely and equally towards the other two ports. As shown in Fig. 3(b) and 3(c), the simulation results for the outgoing wave amplitudes and the turning coefficient are almost the same as the corresponding ones in the T-configuration. The amplitudes of the outgoing waves in the horizontal waveguides (red curve in Fig. 3(b)) drop to a minimum of nearly 1% at 250.8 Hz, while the amplitudes of the outgoing waves (green curve) in the vertical waveguides reach maximum. As indicated in Fig. 3(c), the apparent energy loss in the horizontal waveguides is almost the same as the outgoing wave intensity in the vertical ones. The actual dissipation is about 3 % at the maximum CPC frequency. As is in the T-configuration, the CPC is turned off when the relative phase difference between the two incoming waves is 180°.

The T-configuration in Fig. 1 can be regarded as a special case of the X-configuration with waveguide-3 and -4 bundled together. That is, if one assumes that the waves in waveguide-3 and -4 are always identical, then one can reproduce CPC in the T-configuration. The zero reflection at the T-junction for the side branch is actually a manifestation of the CPC process for the two identical incoming waves in waveguide-3 and -4, respectively. This is the reason why the cross section area of the side branch in the T-configuration must be twice the ones of the main waveguides.

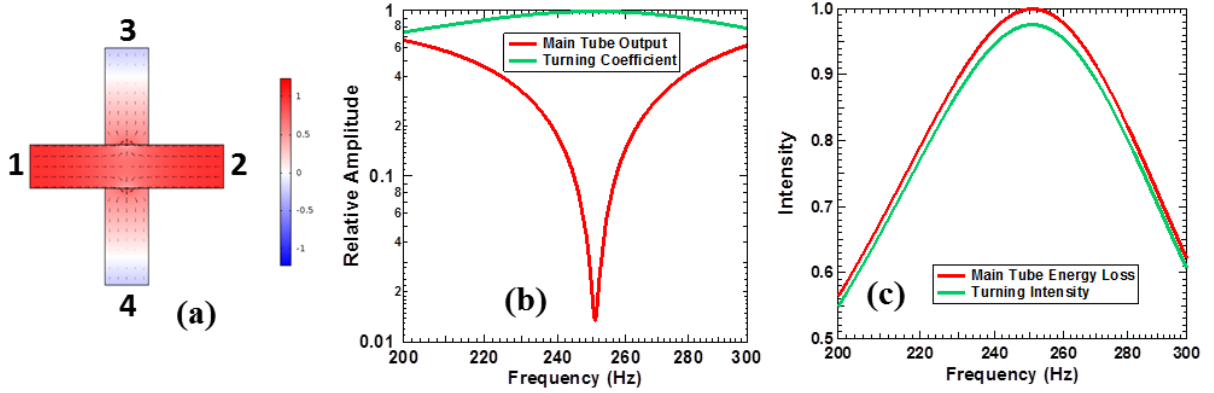

Figure 3. The simulation results of an X-configuration CPC process. (a) The pressure and air velocity fields. A DMR is mounted at the entrances of port-3 and port-4. (b) The amplitude of the outgoing waves (red) and the turning amplitude (green) as a function of frequency. (c) The main waveguide apparent CPA intensity (red) and the turning intensity (green).

There are other possible CPC matrices for the X-configurations. The identification of these matrices could significantly expand the domain of CPC, and provide guidance for the designs of suitable scatterers. The matrix in Eq. (2) can be expressed as $\hat{M}_4 = \begin{pmatrix} \hat{A} & \hat{B} \\ \hat{B} & \hat{A} \end{pmatrix}$, with $\hat{A} = \frac{1}{2}\begin{pmatrix} -1 & 1 \\ 1 & -1 \end{pmatrix}$ and $\hat{B} = \frac{1}{2}\begin{pmatrix} 1 & 1 \\ 1 & 1 \end{pmatrix}$. It is straightforward to show that $\hat{A}\cdot\hat{A} = -\hat{A}, \hat{B}\cdot\hat{B} = \hat{B}$, and $\hat{A}\cdot\hat{B} = \hat{B}\cdot\hat{A} = 0$. Two other possible monopole CPC matrices are $\begin{pmatrix} -\hat{A} & \hat{B} \\ \hat{B} & -\hat{A} \end{pmatrix}$ and $\begin{pmatrix} \hat{A} & \hat{B} \\ \hat{B} & -\hat{A} \end{pmatrix}$. It is straightforward to show that they preserve total wave flux and time reversal symmetry. The first one is similar to that in Eq. (2), in that identical incoming waves from any pairs of the four ports will be channeled completely towards the other two ports.

The second one would produce what we refer to as the 'traffic light effect'. Consider equal inputs from port-1 and -3, $\hat{I} = (1\ 0\ 1\ 0)^T$, the output is $\hat{O} = (0\ 1\ 1\ 0)^T$. The incoming wave in waveguide-3 is totally reflection as if blocked by a hard wall, while the incoming wave in waveguide-1 totally passes through the junction to waveguide-2. This is like the wave in waveguide-3 'sees' a 'red light' at the junction, while the wave in waveguide-1 'sees' a green light and passes unimpededly to waveguide-2. The existence of waveguide-4 does not seem to matter, but its presence actually provides the scattering to channel the waves. In the second case, when the two input waves are in the opposite phase form, $\hat{I} = (-1 \ 0 \ 1 \ 0)^T$, then the outgoing waves are $\hat{O} = (1 \ 0 \ 0 \ -1)^T$. Now the wave in waveguide-1 'sees' a 'red light' and is reflected by a perfectly soft boundary, and the wave in waveguide-3 totally passes through to waveguide-4. Waveguide-2 does not seem to exist, but its presence actually provides the necessary scattering at the junction. Such effect demonstrates that a wave in waveguide-3 or -4 could intersect the waves in waveguide-1 and -2, depending on their relative phase. Likewise, the waves in waveguide-1 and -2 also have similar power in the manipulation of the waves in waveguide-3 and -4.

Another interesting scenario is when waveguide-1 carries two waves of the same amplitude and opposite phase while waveguide-3 carries a wave twice the amplitude, $\hat{I} = (1 \ 0 \ 1 \ 0)^T + (-1 \ 0 \ 1 \ 0)^T$, then the wave in waveguide-1 that is in phase with that in waveguide-3 will pass through to waveguide-2, while the out of phase one is reflected back. Although this seems equivalent to $\hat{I} = (0 \ 0 \ 2 \ 0)^T$ with no wave in waveguide-1, one could make the distinction by using two sub-waveguides for waveguide-1 that combine only near the junction, each sub-waveguide having only half the cross section area as the other three waveguides. This could be difficult for acoustic waves but is readily available in fiber optics.

Possible dipolar CPC matrices are $\begin{pmatrix} \hat{B} & \hat{A} \\ \hat{A} & \hat{B} \end{pmatrix}$, $\begin{pmatrix} \hat{B} & -\hat{A} \\ -\hat{A} & \hat{B} \end{pmatrix}$, and $\begin{pmatrix} \hat{B} & -\hat{A} \\ \hat{A} & \hat{B} \end{pmatrix}$ that completely channel dipolar inputs, which can be shown in a straightforward way that they preserve total wave flux and time reversal symmetry. The first dipole scattering matrix will turn dipolar input $\hat{I} = (1 \ -1 \ 0 \ 0)^T$ into dipolar output $\hat{O} = (0 \ 0 \ -1 \ 1)^T$. The other two have similar effects.

The above scattering matrices derived originally from acoustic wave scatterers could be generalized to other scalar waves, as long as they obey time reversal symmetry and linear superposition principle. As the total flux is conserved, it could ease the difficulties in finding the right scatterers in waves other than acoustic ones, especially for quantum wavefunctions where scattering potentials that conserve particle numbers are much more common than the ones that do not.

The CPC process could significantly extend the coherent perfect manipulations of scalar waves, and spin off to CPX, where 'X' stands for any conceivable wave manipulations. Any CPX process is effectively divided into two steps. The incoming waves are first coherent perfectly channeled to other channel(s), in which they could then be manipulated. If the manipulation is absorption, the process will then be the same as CPA, although real absorption takes place in a much larger space well separated from the CPC region. With the absorbing core removed from the scatterer, more flexibility in design of null detectors becomes possible [7]. The reverse CPC in T-configuration splits one incident wave into two identical waves without the involvement of real lasing process. If cascaded further down $n$-fold, $2^n$ identical waves could be generated with relative ease. The two split waves could also serve as the input for a second CPC, instead of using two conventional sound sources prone to imperfections [7]. Such combination could serve as a high sensitivity interferometer with the incident wave first divided into a perfectly matched pair, and then combined as the inputs for the second CPC for phase sensitive detection.

Total flux conservation implies that the CPC process is the most cascadable in all-wave logic gate operations. It is straightforward to show that the CPC T-configuration can perform all the Boolean operations reported in Ref. 14. The subwavelength nature of the scatterers implies compact and perhaps ultrahigh speed devices if their counterparts in electromagnetic waves and quantum waves could be realized. A network of waveguides having a particular type of scatterer at each junction could have intriguing wave manipulation and logic operation capabilities. The full potential of such networks is yet to be explored.

### Acknowledgement

We sincerely thank P. Sheng and M. Yang for invaluable suggestions. This work was supported by AoE/P-02/12 from the Research Grant Council of the Hong Kong SAR government.

### Authors' contributions

X. Z. and Z. Y. carried out the theoretical investigations. C. M. performed the experiments. Z. Y. wrote most of the manuscript.

### Additional Information

**Competing financial interests:** The authors declare no competing financial interests.

*Corresponding author (phyang@ust.hk)